\documentstyle[epsf,amssymbols,12pt]{article}

\newcommand{\beq}{\begin{equation}}
\newcommand{\eeq}{\end{equation}}        
\newcommand{\bqa}{\begin{eqnarray}}        
\newcommand{\eqa}{\end{eqnarray}}        
\newcommand{\bm}[1]{\mbox{\bf #1}}
\newcommand{\eg}{{\frenchspacing\em e.\hspace{0.4mm}g.{}}}
\newcommand{\ie}{{\frenchspacing\em i.\hspace{0.4mm}e.{}}}
\newcommand{\eq}[1]{{\frenchspacing Eq.~\ref{#1}}}

\title{{\small \hfill WUB 95-13}\\
       {\small \hfill HLRZ 32/95}\\[1cm]
Hyper-Systolic Parallel Computing}
\author{Th. Lippert$^a$, A. Seyfried$^a$, A. Bode$^b$ and K. Schilling$^c$\\[0.2cm]\footnotesize
$^a$Department of Physics, University of Wuppertal, 
    D-42097 Wuppertal, Germany\\[-0.2cm] \footnotesize
    E-mail: lippert@wpts0.physik.uni-wuppertal.de\\ \footnotesize
$^b$Department of Physics, Humboldt University, 
    D-10115 Berlin, Germany\\ \footnotesize
$^c$HLRZ, c/o KFA-J\"ulich, D-52425 J\"ulich,
    Germany and DESY, Hamburg, Germany}

\date{}

\begin{document}
\maketitle

\abstract{A new class of parallel algorithms is introduced
that can achieve a complexity of $O(n^{\frac{3}{2}})$ with respect to
the interprocessor communication, in the exact computation of systems
with pairwise mutual interactions of all elements.  Hitherto,
conventional methods exhibit a communicational complexity of $O(n^2)$.
The amount of computation operations is not altered for the new
algorithm which can be formulated as a kind of $h$-range problem,
known from the mathematical field of Additive Number Theory.  We will
demonstrate the reduction in communicational expense by comparing the
standard-systolic algorithm and the new algorithm on the connection
machine CM5 and the CRAY T3D.  The parallel method can be useful in
various scientific and engineering fields like exact n-body dynamics
with long range forces, polymer chains, protein folding or signal
processing.}

\parindent 0pt
\parskip 12 pt
\section{Introduction}

Within many scientific and engineering applications one is faced with
the intermediate computation of bilocal objects, $f(x_i, x_j)$, on a
given set of numbers $x_i$, $i=1,\dots,n$.  Think for example of the
exact treatment of 2-body forces in n-body molecular dynamics
\cite{SMITH} as employed in astrophysics or thermodynamics of instable
systems, convolutions in signal processing \cite{SIGNAL},
autocorrelations in statistically generated time series
\cite{LIPPERT}, n-point polymer chains with long-range interactions,
protein-folding \cite{LEVITT} or fully coupled chaotic maps
\cite{KONISHI,PALADIN}.

In the case of bilocal objects, symmetric in $i$ and $j$, the entire
computation of all $f(x_i, x_j)$ requires $n(n-1)/2$ combinations of
elements. In the spirit of truly parallel processing, the $O(n^2)$
complexity can in principle be reduced to $O(n)$ per parallel
processor if the number of processors is increased as $n$, the number
of bilocal objects.

On such massively parallel computers with distributed memory, the
elements $x_i$ of the array \bm{x}\ are spread out over many
processing elements, \ie\ local memories.  For this reason such
computations in general require considerable interprocessor
communication. Any progress in reducing the ensuing communication
overhead is therefore welcome\footnote{In molecular dynamics
applications with long range forces, \eg, exact state-of-the-art
calculations are restricted to a number of elements (particles) in the
range of $<100.000$. Exact computations are required near locations
where physical systems exhibit a phase transition.  In such
applications, repeated computations of the elements $f(x_i, x_j)$ have
to be performed with a number of steps in the range of $10^3$ to
$10^5$\label{STATEOF} \cite{GRAPE}.}.

Various methods have been devised in the past to exactly solve the
computational problem; we mention two generic parallel
approaches\footnote{Parallel `linked-cells' algorithms play the major
r\^ole in molecular dynamics with {\em short range} forces
\cite{MULTI}.}: 
\begin{itemize}
\item
The replicated data method \cite{SMITH} deals with $n$ identical
copies\footnote{Here we assume that the numbers of processors, p, and
array elements, n, are equal. The case $p<n$ is discussed in section
\ref{DIFFERENT}.} of the entire array \bm{x}\ that are to be placed
within each processor in advance. On these arrays, the computation is
performed such that each processor $i$ calculates the elements $f(x_i,
x_j)$ for $j=1,\dots,n$.
\item
The parallel organization of the evaluation in form of a systolic array
computation \cite{SMITH,KUNG,PETKOV1,RAINE} proceeds with {\em one} moving
and {\em one} fixed array, that are distributed across the processing
elements.  In each systolic\footnote{In molecular dynamics, systolic array
computations are known as Orrery-algorithm \cite{HILLIS,SCHOLL1}.}
step the moving array is shifted by one position along the processing
elements and subsequently, the computations are performed. After $n-1$ steps,
all pairs $f(x_i, x_j)$ have been generated.
\end{itemize}
Various procedures which combine features of both above mentioned
methods are known in the literature.  Common to all such approaches is that
both the computational complexity and the complexity of the
interprocessor communication is $O(n^2)$.

In this work, we introduce a parallel computing concept that, to a
certain extent, can be regarded as a generalization of systolic array
computations with `pulse' and `circular movement'.  We will point out
that a formulation of the computational problem in the context of
Additive Number Theory---leading to a kind of $h$-range problem
\cite{SCHEID}---defines a new class of algorithms.  Choosing the
base of the $h$-range problem suitably, one can recover the classical
systolic realization of the computational problem.  In addition,
various other bases can be contrived that allow to diminish the
complexity of the interprocessor communication.  We will present a
selection of extremal (shortest) bases and a `regular' base that both
reduce the complexity of the interprocessor communication to
$O(n^\frac{3}{2})$.  The regular base is nearly optimal and well
suited for massively parallel machines.

The new method, further denoted as `hyper-systolic', has in common
with the standard-systolic array computation that only {\em one} array
is moving in each step in a regular communication pattern. Therefore,
it is well suited for SIMD and MIMD machines as is the case with
standard-systolic computations. The distinctive feature as compared to
the standard-systolic concept, however, lies in the fact that storing
of shifted arrays is required\footnote{Complexity in (computer) time
is avoided at the expense of complexity in (storage) space.}, similar
to the replicated data method.  But in contrast to the latter where
the required storage goes with $n^2$, a significant reduction in
interprocessor communication is achieved while only a moderate
additional amount of storage space is needed.  To achieve the optimal
hyper-systolic speed-up, the required storage goes with
$n^{\frac{3}{2}}$.  We will demonstrate that, in view of the
state-of-the-art problem sizes mentioned in footnote~\ref{STATEOF},
our method can be of considerable practical importance.

In section \ref{CP}, we present the generic form of the computational
problem to solve, in section \ref{SYSTOLIC}, we review the
standard-systolic concept.  In order to provide a graphical
illustration, we will deal with systolic automata models
\cite{PETKOV1,PETKOV2}.  A suitable generalization of the latter 
which gives us the means to graphically represent the hyper-systolic
parallel computing structure is worked out in section \ref{HYPER}. In
section \ref{APPLICATIONS}, we discuss implementation issues
concerning the hyper-systolic method on parallel machines and we
compare standard-systolic computations on the Connection Machine CM5
with their hyper-systolic counterpart. Furthermore we present results
from measurements on the Cray T3D.  Finally, we summarize and give an
outlook.

\section{The Computational Problem\label{CP}}

The generic computational task we deal with in the following is the
calculation of
\beq
y_i := \sum_{j=1}^{n} f(x_i,x_j), \qquad i=1,2,3,\dots,n,\qquad i\ne j,
\label{GENERIC}
\eeq
with the input array (sequence)
\beq
\bm{x}=(x_1,x_2,x_3,\dots,x_n)
\eeq
and the resulting array
\beq
\mbox{\bf y}=(y_1,y_2,y_3,\dots,y_n).
\eeq
The number of combinations $\{ x_i,x_j\} $ required for the
computation of all $f(x_i,x_j)$ symmetric in $i$ and $j$ is
\beq
\left( \!\!
\begin{array}{c}n\\2\end{array} \!\! \right) =\frac{n(n-1)}{2},
\eeq 
\ie, the computational complexity inherent to Eq.\ \ref{GENERIC} is
$O(n^2)$.

In molecular dynamics terms, the sequence $\mbox{\bf x}$ can \eg\ be
thought of as the coordinates of $n$ particles, where the sequence
$\mbox{\bf y}$ describes the potential (or equivalently a component of
the force) which particle $\#$ $i$ is subject to in the presence of
all the other particles.

\section{Standard-Systolic Computation\label{SYSTOLIC}}

We first review the systolic realization of Eq.~\ref{GENERIC} from
which our further considerations start.  We use the concept
of systolic automata models to illustrate parallel computing
structures \cite{PETKOV1,PETKOV2}.

\begin{figure}[tb]
\centerline{\epsfxsize=8cm\epsfbox{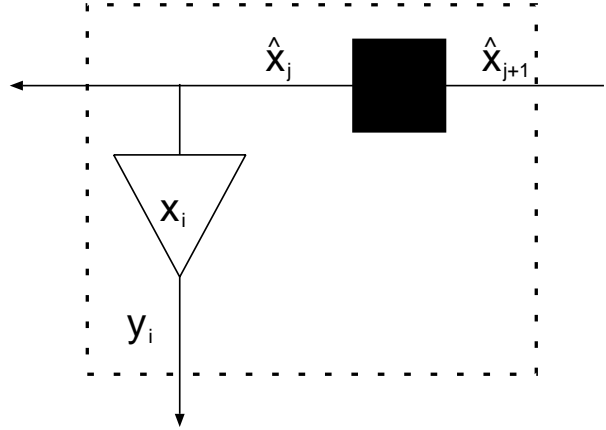}}
\caption[]{
Processing element (open triangle) and delay element (black square),
arranged in a systolic automaton cell (large dotted square).  The
element $x_i$ resides stationary in the processing element.  In each
clock step the delay element delivers another data element $\hat{x}_j$
of the moving array.  The function $f(x_i,x_j)$ is computed and
successively added to $y_i$ that is resident in the processing element
as well.
\label{PEDE}}
\end{figure}

\begin{figure}[htb]
\centerline{\epsfxsize=\textwidth\epsfbox{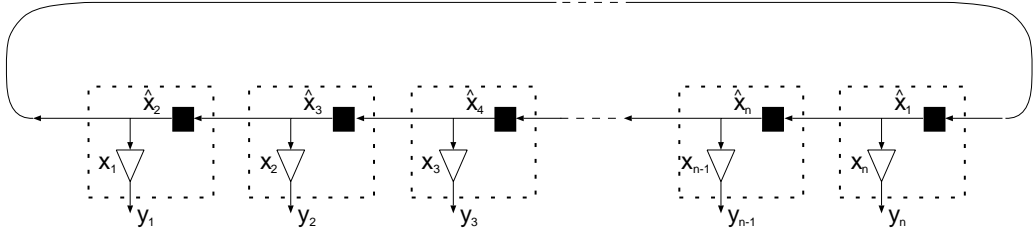}}
\caption[]{
Systolic array of abstract automaton cells with uniform 
next neighbour connections in toroidal topology.
\label{CELL}}
\end{figure}

A systolic automaton consists of cells with data transfer and
processing units, see Fig.~\ref{PEDE}, that are arranged in a regular
grid.  
The cells are coupled to each other in a uniform next-neighbour
wiring pattern as depicted in Fig.~\ref{CELL}.
The processing elements (pe) are drawn as open triangles. They realize
functions of equal load between consecutive communication events.  The
data transfer elements are drawn as black squares. They are delay
elements (de) that represent abstractions for memory locations in
which data is shifted in and out in regular clock steps\footnote{By
`clock step' we denote the clock step of the abstract automaton rather
than a physical computer clock step.}.  Data processing and transfer
are completely pipelined\footnote{ Precise definitions of systolic
arrays and algorithms along with many examples and applications can be
found in two monographs by N.~Petkov \cite{PETKOV1,PETKOV2}}.

\begin{figure}[htb]
\centerline{\epsfxsize=\textwidth\epsfbox{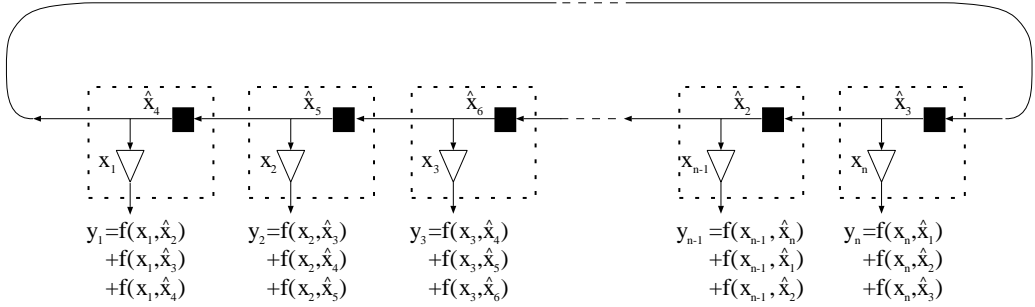}}
\caption[]{
Data location of the systolic array after the third clock step.
\label{SYS3}}
\end{figure}

The graphical structural counterpart of Eq.\ \ref{GENERIC} is given in
Fig.\ \ref{CELL}.  The cells are consecutively arranged in a linear
order.  Each cell is connected with its left and right next
neighbours.  Note that the systolic computation of Eq.\
\ref{GENERIC} leads to a toroidal topology of the linear array.
At clock step $0$, a sequence $\bm{x}$ of $n$ data elements is
distributed over $n$ processing elements. This sequence will stay
fixed in the following process.  Initially a copy $\hat{\bm{x}}$ of
the array $\bm{x}$ is made. The array $\hat{\bm{x}}$ is shifted and
the processing on all cells is performed subsequently clock step by
clock step.  In Fig.\ \ref{CELL}, the situation is shown for the first
clock step.  Fig.\ \ref{SYS3} illustrates the state of the systolic
computation after 3 clock steps.

The small black boxes stand for the de's, the function of which is to
shift the data element $\hat{x}_i$ in one clock step from cell $\#$
$i$ to cell $\#$ $(i+1)$.  Subsequently, the pe's perform the
numerical computation of the function $f$, \ie, they link the elements
of the fixed array $\bm{x}$ with the elements of the shifted array
$\hat{\bm{x}}$. The result in the $i$-th cell is added to the
resulting data element $y_i$ of the array $\bm{y}$ such that after
$n-1$ steps, the resulting data elements are distributed over all
processing elements\footnote{In terms of the molecular dynamics
example, each particle coordinate is located together with its
respective force value in the same processing element.}.

The mapping of the abstract automaton model onto a real parallel
computer is straightforward: the pe's are identified with the
processors of the parallel machine, the de's symbolize registers or
memory locations exchanging data via the interprocessor communication
network. Of course, for the parallel machine a communication network
is required that allows to realize the above considered particular
communication pattern with a toroidal topology efficiently.  This is
the case for the most SIMD and MIMD parallel machines, see the
implementation of the hyper-systolic algorithm described in section
\ref{APPLICATIONS}.

The complexity of the systolic computation of Eq.\ \ref{GENERIC} is of
order $n^2$ for both the data transfer operations and the processing
operations, \ie\ exactly $n(n-1)$ processing operations and communication
events were required if the method would be applied to the full array with
$n$ elements.  One can improve on the scaling of the systolic realization
taking into account the symmetries.  The computations can be reduced to
$n(n-1)/2$ processing operations but now $n(n+1)$ communication events have
to be performed, since also the result array has to be shifted $n/2$ times
and finally one reverse shift must be carried out.  This method, called
`Half-Orrery' algorithm, has been introduced in a recent paper by Scholl
and Alimi \cite{SCHOLL1}. The limiting behaviour of $O(n^2)$, however, is
the same as before, for both communication and CPU operations.

\section{The Hyper-Systolic Algorithm\label{HYPER}}

The main message of this paper is that the total time expenditure for
the interprocessor communication in the computation of Eq.\
\ref{GENERIC} can be reduced considerably as it is possible to achieve
a complexity of $O(n^{\frac{3}{2}})$ by a suitable re-organization of the
data-movement for the communicational part of Eq.\ \ref{GENERIC}, at a
moderate expense of storage space. The amount of processing operations is
not altered.

\subsection{Motivation\label{MOTIV}}

Consider the matrix $C$ which explicitly enumerates the data elements of
the shift-array $\hat{\bm{x}}_t$ for the clock steps $t=0,1,2,\dots,n-1$,
as delivered successively in the systolic implementation of section
\ref{SYSTOLIC}:
\begin{equation}
C= \left(
\begin{array}{lccccccc}
        &1  &2&3&\ldots& &   &n    \\
        &n  &1&2&\ldots& &n-2&n-1  \\
        &n-1&n&1&\ldots& &n-3&n-2  \\
        &.  & & &      & &   &.    \\      
        &.  & & &      & &   &.    \\      
        &.  & & &      & &   &.    \\      
        &2  &3&4&\ldots& &n  &1
\end{array}
\right)
\begin{array}{l}
        $t=0$      \\
        $t=1$      \\
        $t=2$      \\
        $t=3$      \\      
                   \\      
                   \\      
        $t=n-1$.
\end{array} 
\label{SHIFT}
\end{equation}
We observe in (\ref{SHIFT}) that all combinations of different
elements actually occur $n$ times, if every shifted array
$\hat{\bm{x}}_t$ is stored for $t=0,1,2,\dots,n-1$ and combinations
between all the rows are allowed.  Note that storing the shifted
arrays is in contrast to the systolic concept where only {\em one}
fixed and {\em one} moving array came into play.  Here, as time
proceeds, an increasing number of `resident' arrays is needed, with
only one array in movement.

The matrix $C$ shows that an $n$-fold redundancy of pairings is encountered
if all shifted arrays are stored.  It appears natural to reduce this
redundancy of pairings and optimize by storing a few arrays
$\hat{\bm{x}}_{t}$ only, for a suitably selected subset of time steps $t$.

To illustrate the idea, let us consider a simple example for $n=16$; from
matrix $C$ in Eq.\ \ref{SHIFT} we take five rows:
\begin{equation}
C'= \left(
\arraycolsep=1.5pt
\begin{array}{llllllllllllllll}
        1 &2 &3 &4 &5 &6 &7 &8 &9 &10&11&12&13&14&15&16        \\
        16&1 &2 &3 &4 &5 &6 &7 &8 &9 &10&11&12&13&14&15        \\  
        14&15&16&1 &2 &3 &4 &5 &6 &7 &8 &9 &10&11&12&13        \\       
        12&13&14&15&16&1 &2 &3 &4 &5 &6 &7 &8 &9 &10&11        \\          
        8 &9 &10&11&12&13&14&15&16&1 &2 &3 &4 &5 &6 &7         \\   
\end{array}
\right).
\label{SMALLEX}
\end{equation}
It suffices to discuss in $C'$ combinations with element $\#$ $1$ only,
because of the toroidal topology of the systolic array.  In the first
column the combinations $\{ 1,16\} $, $\{ 1,14\} $, $\{ 1,12\} $ and $\{
1,8\} $ can be found, in the second column $\{ 1,2\} $, $\{ 1,15\} $, $\{
1,13\} $ and $\{ 1,9\} $ occur, in the fourth column we have $\{ 1,3\} $,
$\{ 1,4\} $ and $\{ 1,11\} $, in the sixth column we have $\{ 1,5\} $ and
$\{ 1,6\} $ and in the tenth column the yet missing combinations $\{ 1,10\}
$ and $\{ 1,7\} $ are present.

Thus, in order to generate all pairs for $n=16$ only $5$ stored arrays
are required. 

We note further that to realize Eq.\ \ref{GENERIC}, we cannot just add
the outcome of the computations to one resulting array: for every row
$\hat{\bm{x}}_{t}$ to be stored, a separate resulting array
$\hat{\bm{y}}_{t}$ is required. These arrays have to be shifted back
in steps inverse to the foregoing. Then they can be added to give the
final resulting array $\bm{y}$.

Therefore, for this small example, in total $8$ shift-operations are
required. In the standard systolic process $15$ shifts of the moving
array must be carried out.  Using the symmetry in $i$ and $j$, in
total, $17$ shifts are to be performed (Half-Orrery algorithm
\cite{SCHOLL1}), because in this case the result array must be transported
too, and finally shifted back.  The array length $4$ represents the
cross-over between standard-systolic and the new way to compute
\eq{GENERIC} concerning the number of array shifts.  Going to larger
arrays, the new algorithm becomes more and more advantageous.

\subsection{The algorithm\label{HYPERAL}}

We now give a general prescription to compute Eq.\ \ref{GENERIC}
according to the ideas exposed in the last section.  The new algorithm
is called {\em hyper-systolic}, the name referring to the topological
features of the underlying communicational structure:
\begin{enumerate}
\item
For a given array $\bm{x}$ of length $n$, $k$ copies
$\hat{\bm{x}}_{t}$ are generated by shifting the original array
$\bm{x}$ $k$ times by a stride $a_t$ and storing the resulting arrays as
$\hat{\bm{x}}_{t}$, $1\le t\le k$.  The variable stride $a_t$ 
of the shifts must be chosen such that
\begin{enumerate}
\item
all pairings of data elements occur at least once, 
\item
the number of equal pairings  is minimized,
\item
the number of shifts $k$ is minimized,
\item
the number of different strides $a_t$ is minimized.
\end{enumerate}
\item
The required $n(n-1)/2$ results $y_i$ according to Eq.\ \ref{GENERIC}
are successively computed and are added to $k$ resulting arrays
$\hat{\bm{y}}_{t}$.
\item
Finally, applying the inverse shift sequence, the arrays
$\hat{\bm{y}}_{t}$ are shifted back and 
corresponding entries are summed up to build the 
elements of the final array $\bm{y}$.
\end{enumerate}

In the above defined hyper-systolic parallel computing structure, in
general, $2k$ intermediate arrays are required. Later, we will come to
special realizations that work with less intermediate arrays.

\subsection{Graphical representation\label{GRAPHICAL}}

\begin{figure}[htb]
\centerline{\epsfxsize=8cm\epsfbox{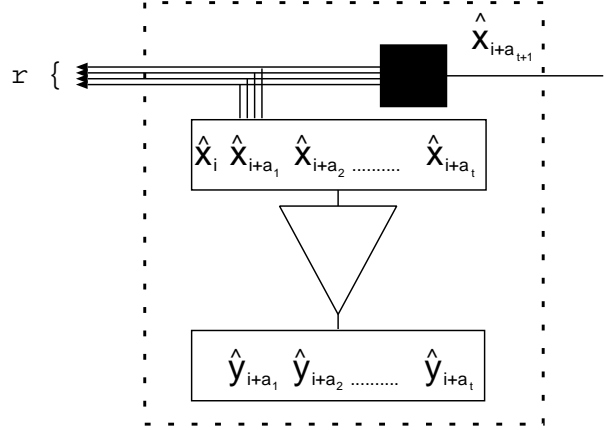}}
\caption[]{
Processing element (open triangle), delay element (black square) and
store elements (se) (open squares), arranged in a hyper-systolic
automaton cell (large dotted square).  The element $\hat x_i$ resides
stationary in the storing element.  First, in each clock step $t$, the
delay element delivers a new data element $\hat{x}_{i+a_t}$ that is
both stored in the local se and transmitted further. Subsequently the
computing is performed on the relevant combinations of yet stored
elements. The results are stored in the corresponding memory locations
$\hat{y}_{i+a_t}$. Note that each de is equipped with $r$ connectors
to account for the $r$ different strides $a_t$. Second, the results
$\hat{y}_{i+a_t}$ have to be successively shifted back, using a stride
sequence inverse to the above, and added up to yield the final result.
\label{PEDE:HYP}}
\end{figure}

The concept of systolic automata models can be transferred to the
hyper-systolic parallel computing structure.  For the graphical
representation we use the objects introduced in section
\ref{SYSTOLIC}.  Fig.\ \ref{PEDE:HYP} shows the $i$-th  hyper-systolic cell
unit. 
The de's again represent input and output.  Additionally we use open
rectangles to symbolize the memory locations for the stored data
elements $\hat{{x}}_{i+a_t}$ and $\hat{{y}}_{i+a_t}$, $1\le t\le k$,
$1\le i \le n$.  Thus we account for the original systolic
functionality of the de's acting on transient data elements
only\footnote{We note that the index $i+a_t$ is to be taken as $i+a_t
-n$ for $i+a_t>n$ (or $\mbox{mod}(i+a_t-1,n)+1$) because of the
toroidal topology of the hyper-systolic array.}.  The de's now possess
$1$ input and $r$ output connectors, one separate connector for each
required shift width $a_t$.  In general, the number of connectors $r$
is not equal to $k$.  It depends on the chosen hyper-systolic scheme.

\begin{figure}[htb]
\centerline{\epsfxsize=\textwidth\epsfbox{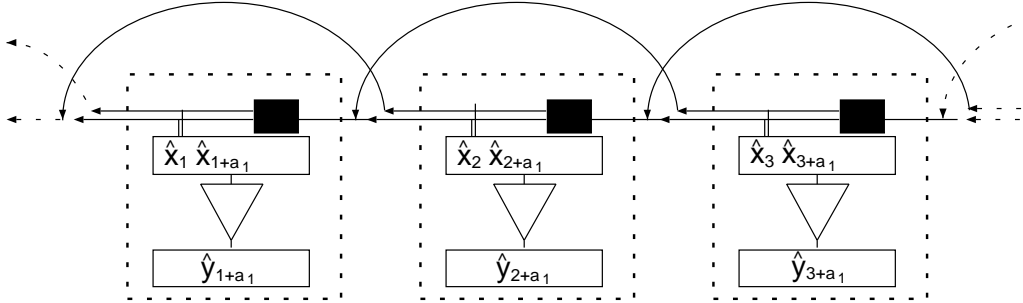}}
\caption[]{
3 abstract automaton cells as part of a hyper-systolic array with $r=2$
uniform next neighbour connections and toroidal topology. In this
simplified case, the two strides are of length 1 and 2.
\label{CELL:HYP}}
\end{figure}

Again the cells are arranged in a uniform grid; in addition to the
next-neighbour wiring, the new connections between the processors are
drawn in Fig.\ \ref{CELL:HYP}.
They correspond to the shifts of data elements by strides $a_t$. According
to the number $r$ of different strides $a_t$, $r$ direct connections would
be optimal.  From the point of view of the original one-dimensional
systolic array structure---due to these additional communication
connections---the (one-dimensional) space of processing elements has
acquired topologically non-trivial interconnections.  Fast interprocessor
data transfer to distant cells can be achieved in one clock step via these
interconnections that are used as shortcuts.  Thus the characterization of
our new algorithm as {\em hyper-systolic}.

The pe's now don't realize functions of equal load between consecutive
communication events. The amount of processing operations $g(t)$
increases with an increasing number $t$ of stored arrays
$\hat{\bm{y}}_{t'}$, $1 \le t'\le t$.  We emphasize again that
hyper-systolic data movement and storing extends the standard-systolic
concept.

\begin{figure}[htb]
\centerline{\epsfxsize=\textwidth\epsfbox{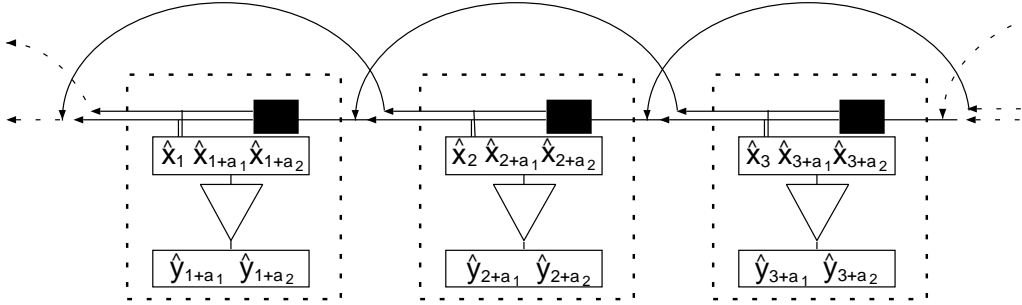}}
\caption[]{
Data location on a part of the hyper-systolic array after the second clock
step.
\label{HYPSYS2}}
\end{figure}

Fig.\ \ref{CELL:HYP} indicates the graphical structural counterpart of
Eq.\ \ref{GENERIC} for the hyper-systolic algorithm; we have plotted
the situation for $t=1$.  At clock step $t=0$, a sequence $\bm{x}$ of
$n$ data elements is distributed over $n$ processing elements.  In
every clock step $t=0,1,2,\dots,k$, a copy $\hat{\bm{x}}_{t}$ of the
array $\hat{\bm{x}}_{t-1}$ is generated and, in the next clock step,
the array is shifted by a stride $a_t$.  Subsequently, the new array
$\hat{\bm{x}}_{t}$ is stored in the se's.  Then the processing can be
done by the pe's involving ever more of the stored arrays for
increasing $t$.  Fig.\ \ref{HYPSYS2} shows the situation after 2 clock
steps. Note however that only {\em one} array is moving as is the
case in the standard-systolic computation.

In clock step $t$, the $g(t)$ results in the $i$-th cell are added to
the corresponding elements ${y}_{i+a_{t'}}$, $1\le t'\le t$.  Finally,
when $n(n-1)/2$ processing operations have been performed, the $k$
resulting arrays $\hat{\bm{y}}_{t}$ are shifted back by strides
inverse to the above sequence and successively added up to the result
array $\bm{y}$.

\subsection{Additive Number Theory and hyper-sys\-tol\-ic parallel 
            processing}

As explained in section \ref{HYPERAL}, our task is to construct a
sequence of strides $A_k=\{ a_0=0,a_1,a_2,a_3,\dots,a_k\} $ that---in 
the course of the
hyper-systolic process---provides us with $k+1$ sequences
$\hat{\bm{x}}_{t}$, $t=0,1,2,\dots,k$, represented as a rectangular
matrix $C'$\footnote{Note that we have introduced the
zero-element.}:
\begin{equation}
C'= \left(
\begin{array}{ccccccc}
        x_{1+a_{0}}&x_{2+a_{0}}&x_{3+a_{0}}&\ldots& & &x_{n+a_{0}}        \\
        x_{1+a_{1}}&x_{2+a_{1}}&x_{3+a_{1}}&\ldots& & &x_{n+a_{1}}        \\              
        .& & &      & & &.        \\      
        .& & &      & & &.        \\      
        .& & &      & & &.        \\      
        x_{1+a_{k}}&x_{2+a_{k}}&x_{3+a_{k}}&\ldots& & &x_{n+a_{k}}        \\              
\end{array}
\right).
\label{RECTANGULAR}
\end{equation}
As mentioned above, in $C'$ all indices are understood as
$\mbox{mod}(i+a_1-1,n)+1$.  It is required that all combinations of
elements from $1$ to $n$ can be constructed along the columns of $C'$.
The integer $k$ is to be chosen as small as possible. $a_t$ is the
stride by which $\hat{\bm{x}}_{t-1}$ is cyclically shifted to render
$\hat{\bm{x}}_{t}$.

We formulate the algorithmic problem in terms of Additive Number
Theory:

Let $I$ be the set of integers $m=\{ 0,1,2\dots,n-1\}\in \Bbb{N}^n_0,
n\in \Bbb{N}$. Find the ordered set $A_k=(a_0=0,a_1,a_2,a_3,\dots,a_k)\in
\Bbb{N}^{k+1}_0$ of $k+1$ integers, with $k$ minimal, 
where each $m\in I$, ($0\le m\le n-1$), can be represented at least
once as the ordered partial sum
\beq
m=a_i+a_{i+1}+\dots +a_{i+j} 
\label{ADD}
\eeq
or as 
\beq
m=n-(a_i+a_{i+1}+\dots +a_{i+j}), 
\label{ADDI}
\eeq
with 
\beq 0\le
i+j\le k,\qquad i,j \in \Bbb{N}_0.  
\label{ADDII}
\eeq
This formulation with the extremal base $A_k$ is reminiscent of the
`postage stamp problem', a famous problem in Additive Number Theo\-ry
\cite{SCHEID,DJAWADI}.  In contrast to the original postage stamp
problem, here we deal with sequences $A_k$ instead of sets and with
ordered partial sums. In these terms we are looking for the $h$-range
$n(h,A_h)$ of the extremal basis $A_h$ with only ordered partial sums
of the elements of $A_h$ allowed as given by Eqs.\ \ref{ADD},
\ref{ADDI} and \ref{ADDII}.

Unfortunately, our problem is just as little solvable in general as
the postage stamp problem.  Nevertheless, later we will propose an
approximate solution that carries optimal asymptotic scaling
properties {\it and} is very well suited for the hyper-systolic
realization on parallel SIMD and MIMD machines.

The formulation in terms of an $h$-range problem appears to provide a
quite general framework covering a whole class of algorithms: any
chosen base $A$ induces its specific new algorithm.

The realization of $A_k$ with $k=n-1$ brings us back to the
standard-systolic algorithm with the base
\begin{equation}
        A_{n-1}=\{0,a_1,\ldots,a_{n-1}\}, 
        \qquad a_i=1\qquad\forall\,\, i=1,\dots,n-1,
\end{equation}
that features a communicational complexity of $O(n^2)$.

The realization 
\begin{equation}
        A_{\frac{n}{2}}=\{0,a_1,\ldots,a_{\frac{n}{2}}\}, 
        \qquad a_i=1\qquad\forall\,\, i=1,\dots,\frac{n}{2},
\end{equation}
leads to the Half-Orrery algorithm if the result array is moved too.

A lower bound on the minimal possible $k$, with optimal complexity,
can be derived by the following consideration:

The total number of combinations required is $n(n-1)/2$. Let the
matrix $C'$ of Eq.~\ref{RECTANGULAR} be realized by $k$ shifts, \ie\
by $k+1$ arrays.  The number of possible combinations between the
elements of a given column of $C'$ follows as $\left( \!\!
\begin{array}{c}k+1\\2\end{array} \!\! \right)$.  Thus, the
following unequality holds:
\beq
\frac{n(n-1)}{2}\le
\left( \!\! \begin{array}{c}k+1\\2\end{array} \!\! \right)n,
\eeq
therefore,
\beq
k\geq-\frac{1}{2}+\sqrt{n-\frac{3}{4}}.
\eeq
with $k$ integer.  Obviously, a sequence $A_k$ with less than
$k_{min}$ elements cannot solve the h-range problem presented above.
In other words, the complexity for the interprocessor communication of
any hyper-systolic algorithm can at best be $n\sqrt{n}$.

\subsection{Regular bases\label{REGULAR}}

In choosing a suitable base for the hyper-systolic algorithm, we
better make use of the fastest links of our parallel architecture.  So
we have to meet a constraint: in general, from each processor there is
a limited number $r$ of fast connections to $r$ other processors. This
is, however, very much dependent on the particular parallel machine.
On (hyper)cubic interconnection networks, \eg, $r$ relates to the
dimension of the hypercube. We should prefer shift sequences that are
restricted to exploit the $r$ fast interprocessor connections
available, in the actual implementation of the hyper-systolic method
on a parallel machine.  To a certain extent, the network can provide
shortcuts or `wormholes' through higher dimensions that allow to
accelerate the standard-systolic movement along the one-dimensional
array.

In the case of the standard-systolic array, $r=1$, because only next
neighbour wiring is requested in the systolic process.  This amounts
to take local trains only for long-distance traveling.

The next simple step would be a mix of intercity and local connections
for the long-distance journey. In fact, a nearly optimal choice of a
suitable sequence, leading to a very efficient implementation of the
hyper-systolic ideas and requiring $r=2$ interprocessor connections
only, is given by:
\beq 
\arraycolsep=0pt
\begin{array}{ccccccccc}
A_{k=2K-1} &=& \Big( & 1,1,\dots,1                 &,&\, K,K,\dots,K                  & \Big) &,&\qquad K=\sqrt{\frac{n}{2}},\\ 
           & &       & \underbrace{\makebox[1.6cm]{}}& & \underbrace{\makebox[1.6cm]{}} &       & &                            \\
           & &       & K                           & & K-1 \\
\end{array}
\eeq 
with $K$ shifts by $1$ and $K-1$ shifts by $K$.  The length of the base 
is $k=2K+1$. It can easily be
shown that such a base leads to an h-range of $n$ with respect to
Eqs.~\ref{ADD} and \ref{ADDI}.  The communicational complexity of the
hyper-systolic method using the regular shift sequence with shift
constant $K$, \ie\ strides $a_t=1$ for $t\le K$ and $a_t=K$ for $t>K$,
turns out to be $2n(2\sqrt{n/2}-1)$.  Note that we have included an
additional factor $2$.  This is due to the fact that the number of
required reverse shifts the result arrays are subject to also is given
by $2\sqrt{n/2}-1$.

In order to illustrate the hyper-systolic array generated  
by the regular base let us
plot an example where $n=32$ and $K=\sqrt{\frac{n}{2}}=4$ in Fig.~\ref{M32}:
\begin{figure}[htb]
\centerline{\epsfxsize=\textwidth\epsfbox[56 697 589 772]{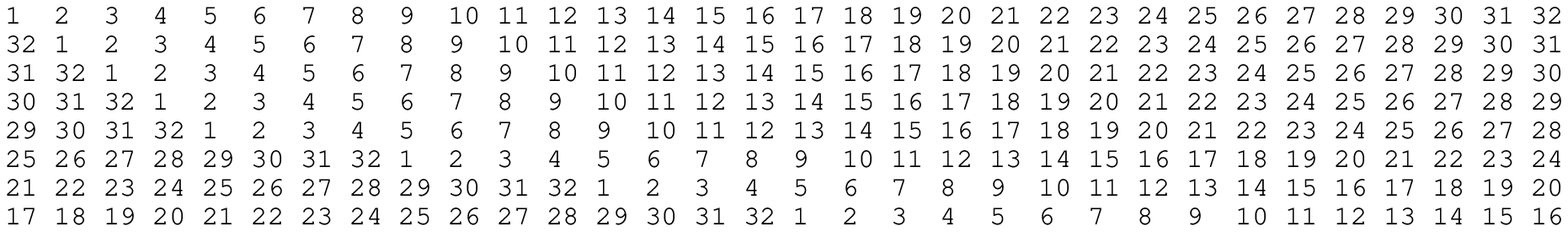}}
\caption[dummy]{Matrix C' for $n=32$ using the regular base
$A_7=(1\, 1\, 1\, 1\, 4\, 4\, 4\,)$.\label{M32}}
\end{figure}

Building combinations along the columns, all pairings of the numbers
$1$ to $32$ between the different rows can be found at least once.
Note that the last row requires a separate treatment as, only in
that case, the pairs occur twice.

Obviously, $K=\sqrt{\frac{n}{2}}$ makes only sense for integer values
$K$.  Therefore, $\frac{n}{2}$ should be a square number.  However, if
this is not possible, one can switch over to a suitable neighbouring
integer value $K \gtrless \sqrt{\frac{n}{2}}$ as shift constant.

\subsection{Shortest bases\label{Shortest}}

The regular bases just introduced are well suited for massively-parallel
machines with a large number of processors.

Let us consider the ratio $R$ between standard- and hyper-systolic
communicational complexity, the ``gain-factor'',
\beq
R=\frac{n+1}{2(2\sqrt{\frac{n}{2}}-1)}\approx \sqrt{\frac{n}{8}}.
\label{RATIO}
\eeq
On a machine with 32 processors, \eg, we can gain a factor of $2$ in
interprocessor communication events between standard-systolic and
hyper-systolic computations. This ratio is increasing rapidly going to
machines with larger numbers of processors and \eg\ becomes
$11$ for 1024 processors.

On coarser grained computers with $\#$ of processors $\le 64$, other
bases than the regular type might be more advantageous.  

In absence of analytical solutions, for various numbers of processors
$p$ from 4 to 64 we have constructed well suited extremal bases by
straightforward computation. We tried to prefer bases with elements
being a power of 2, as these strides might lead to fastest
interprocessor communication on most parallel machine's networks.

Table~\ref{OPTIMAL} collects our selection of extremal bases.
\begin{table}[htb]
\begin{center}
\begin{tabular}{| l l l | l l l  |}
\hline
$p$   &   $k$   & $A_k$ & $p$   &   $k$   & $A_k$ \\
\hline
4   &   2   & 1 1 & 5   &   2   & 1 1 \\
6   &   2   & 1 2 & 7   &   2   & 1 2 \\
8   &   3   & 1 1 2 & 9   &   3   & 1 1 2\\
10  &   3   & 1 2 2 & 11  &   3   & 1 2 2\\
12  &   3   & 1 2 4 & 13  &   3   & 2 1 4\\
14  &   4   & 1 1 1 4 & 15   &   4   & 1 1 1 4\\
16  &   4   & 1 2 2 4 & 17   &   4   & 1 1 2 8\\
18  &   4   & 2 1 8 4 & 19   &   4   & 3 6 8 4\\
20  &   5   & 1 1 1 4 4 & 21   &   4   & 3 10 2 5\\
22  &   5   & 1 1 1 4 4 & 23   &   5   & 1 1 1 4 4\\
24  &   5   & 1 1 1 4 8 & 25   &   5   & 1 1 2 8 8\\
26  &   5   & 1 2 3 4 4 & 27   &   5   & 1 2 2 8 8\\
28  &   6   & 1 1 1 4 4 4 & 29   &   6   & 1 1 1 4 4 4\\
30  &   6   & 1 1 1 4 4 4 & 31   &   6   & 1 1 1 4 4 4\\
32  &   6   & 1 1 1 4 4 8 & 36   &   6   & 3 6 1 8 4 12\\
48  &   7   & 3 1 11 7 16 4 4 & 64   &   8   & 1 1 12 3 10 8 20 4\\
\hline
\end{tabular}
\end{center}
\caption[dummy]{Shortest bases for various machine sizes.\label{OPTIMAL}}
\end{table}
On a 32 node parallel machine, \eg, we can theoretically gain a factor
$R=2.75$ compared to $R=2$ using the regular base.

For $p>32$ it becomes increasingly difficult to find extremal bases
with elements near a power of $2$ that would be shorter than the
regular base. For $p=64$, however, if we do not require all elements
being a power of $2$, the shortest base is only $k=8$ elements in length
whereas the regular base needs $k=11$ elements.

Again we illustrate the hyper-systolic algorithm using an extremal
base for $n=32$, see Fig.~\ref{E32}:
\begin{figure}[htb]
\centerline{\epsfxsize=\textwidth\epsfbox[56 697 589 772]{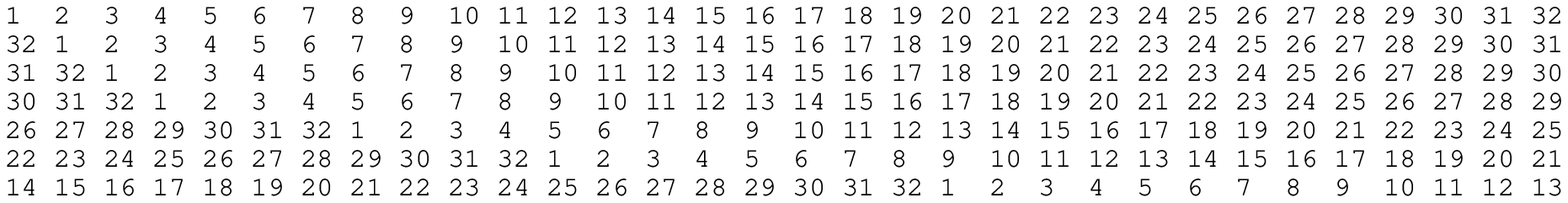}}
\caption[dummy]{Matrix C' for $n=32$ using the extremal base
$A_6=(1\, 1\, 1\, 4\, 4\, 8\, )$.\label{E32}}
\end{figure}

\subsection{Coarse grained limit\label{DIFFERENT}} 

The number of processors $p$ of a given parallel machine and the
number of array elements $n$ can be different.  In order to adapt both
the standard-systolic and the hyper-systolic concept to this
situation, we can partition the array of $n$ elements into
$\frac{n}{p}$ subarrays, such that each processor deals with
$\frac{n}{p}$ data elements.  Each subarray (of $p$ elements) is
distributed across the $p$ processors and can be treated by the
standard-systolic or hyper-systolic method such that successively all
required combinations between all $n$ data elements can be
constructed.

\begin{figure}[htb]
\centerline{\epsfxsize=11.6cm\epsfbox[54 666 318 770]{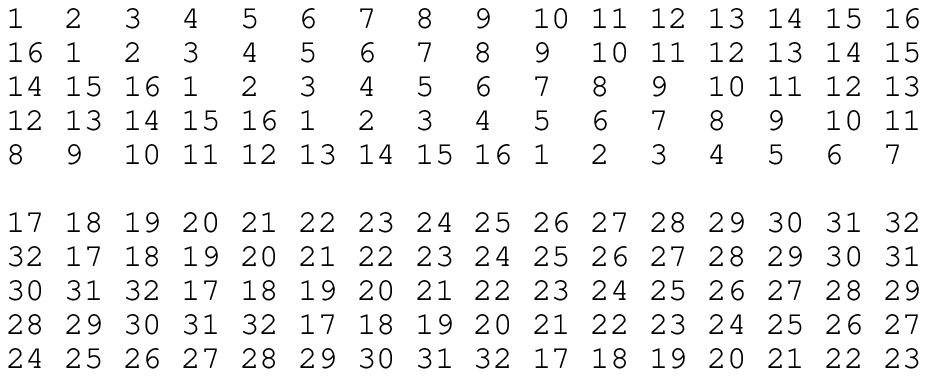}}
\caption[dummy]{Two matrices C' for $n=32$ and $p=16$ using the extremal base
$A_4=(1\, 2\, 2\, 4)$.\label{PGREATERN}}
\end{figure}

To illustrate the situation, the two stored hyper-systolic arrays for
$p=16$ and $n=32$ using the extremal base of Table~\ref{OPTIMAL} are
depicted in Fig.~\ref{PGREATERN}.

Considering the complexity, in the standard-systolic (Half-Orrery)
computation, $n(p+1)$ interprocessor data movements have to be carried
out, whereas in the hyper-systolic case the number of interprocessor
movements follows as $2n(2\sqrt{p/2}-1)$.  The improvement again can
be described by the ratio $R$ of \eq{RATIO},
\beq
R\approx \sqrt{\frac{p}{8}}.
\eeq

With $n \gg p$ it becomes important to compare the pure computation
time vs.\ the interprocessor communication time as the former scales
as $O(n^2)$. If we take $n$ very large the parallel machine turns back
to a scalar machine, \ie, interprocessor communication becomes
irrelevant for both standard-systolic and hyper-systolic computations.
Of course there is a limit in cpu-power, and the latter determines the
maximal possible $n$ that can be computed in reasonable time.  In
practice, for that value of $n$, one has to determine the amount of
interprocessor communication overhead compared to the pure computation
in order to asses a substantial improvement of the hyper-systolic
method vs.\ the standard-systolic method.

\section{Performance Tests and Results\label{APPLICATIONS}}

We will demonstrate in the following section that the hyper-systolic
method, applied to the computation of Eq.\ \ref{GENERIC} on massively
parallel computers, is indeed superior to standard-systolic
implementations in real life applications. In the realistic setting of
an exact molecular dynamics simulation with long range forces, we
study the performance of standard- hyper-systolic n-body dynamics with
mutual gravitational interactions on connection machines CM5 with
numbers of nodes in the range of $16$ to $1024$ and $320$-node Cray T3D.

\subsection{Molecular dynamics with long-range for\-ces}

The gravitational force $\vec F(\vec x_i)$ acting on the $i$-th
particle within an ensemble of $n$ particles of equal mass $m=1$
reads: \beq \vec F(\vec x_i) = \sum_{j=1}^n \frac{\vec x_i - \vec x_j}
                                     {|\vec x_i - \vec x_j|^3}, \qquad
j\ne i, \label{NEWTON} \eeq where the gravitation constant is set to
$1$.  The $\vec f(\vec x_i,\vec x_j)=\frac{\vec x_i - \vec x_j}{|\vec
x_i - \vec x_j|^3}$ by \eq{NEWTON} belong to the class of symmetric
bilocal objects used in \eq{GENERIC}. We deal with two dimensions,
\ie, $\vec x_i = \left({x^1_i\atop x^2_i}\right)$.

In a molecular dynamics simulation, given a distribution of particles
to which coordinates and momenta are assigned at time step $T=l$, the
long range forces between the particles are calculated with respect to
\eq{NEWTON}, and subsequently, the new positions and momenta at time
step $T=l+1$ are found using a suitable integration method like, \eg,
the {\em leap frog} scheme\footnote{The integration is a purely local
procedure, therefore no interprocessor communication is required in
that step.} \cite{LEAPFROG}.  The time-consuming part of the
simulation is the computation of \eq{NEWTON}. In order to improve on
the interprocessor communication we have to concentrate on the
efficient parallel implementation of the latter.  As we here deal with
two dimensions, two arrays of coordinates,
$\vec{\bm{x}}=\left({\bm{x}^1\atop \bm{x}^2}\right)$, describe the
position of the given particle. The momenta do not enter the
computation of \eq{NEWTON}.

\subsection{Implementation issues}    

At this stage the individual features of the programming languages and
network topologies are discussed.

\paragraph{CM-FORTRAN}\cite{CM-FORTRAN}  This parallel
language supports the concept of `virtual' processors.  From the
user's point of view, it allows to emulate a virtual parallel machine
with $n$ virtual processors, whereas the real machine has $p$ real
processors. In principle, the user can write programs as if there
would be a machine with $n$-processors at hand, and the compiler and
the operating system take care of the additional administrative
overhead. CM-FORTRAN appears to be well suited for the straightforward
implementation of a one-dimensional standard-systolic array. The
hyper-systolic method, however, requires explicit control over the
data distribution across the $p$ real processors. Therefore, on a
machine with $p$ real processors, we cut the array into $n/p$ parts
and declare each array of $p$ elements as data-parallel array.

The interprocessor communication network on a connection machine CM5
is based on the fat-tree concept \cite{CM5FATTREE}.  We expect that
communication times to nodes at a `distance' of more than one
processor scale logarithmically with the number of processors.  A
connection machine CM5 meets the requirements of the hyper-systolic
concept with regular array shifts: for any available machine sizes
ranging from 16 to 1024 processors, the shift of a data element to a
location at distance $a_t=K=\sqrt{p/2}$ in the hyper-systolic array,
is achieved nearly as fast as the shift to the next array element
($a_t=1$) because the latency dominates the communication time.

\paragraph{T3D} A message passing parallel programming language, as \eg\ PVM
implemented on the Cray T3D, does not support virtuality.  We set up
the array of $n$ data elements on $p$ processors available in the
straightforward way discussed above, cutting the array into $n/p$
parts.  Note that PVM is not the only way to program message passing
on the T3D.  We did, however, not employ the fast low-level
communication routines that are available.

The T3D's communication network consists of a three-dimen\-sio\-nal
torus.  Unlike the CM5's network, colliding messages are possible on
this network. If the T3D application is run on a small partition it
cannot use the toroidal structure.  However, we expect the latency for
exchange of interprocessor messages to dominate the `node-hopping'
time and therefore only little influence of this effect should be
observable.

\subsection{Performance results on CM5 and T3D}

On the CM5, we want to avoid any interference of the `VP-manage\-ment'
with improvements due to the hyper-systolic computation.  For our
tests, we decided to work with $n=p$ on CM5 and CRAY T3D, \mbox{\ie},
the VP-ratio is $1$ on the CM5.  Therefore, any gain in performance
should theoretically scale as given by
\eq{RATIO}\label{PIPELINING}.\footnote{\label{FOOT:PIPE} We emphasize
that with $n=p$, we don't have the advantage of vectorizing
capabilities or communication pipelining.}

We had access to the connection machine CM5 at GMD/HLRZ, Germany, with
partitions of $16$, $32$, and $64$ nodes and the CM5 at Los Alamos
National Laboratories, USA, with 128, 512 and 1024 node partitions.
Furthermore, we were able to conduct performance measurements on the
320-node Cray T3D at Edinburgh Parallel Computing Centre (EPCC) where
we tested our algorithm using partitions of $16$, $32$, $64$, $128$
and $256$ nodes.

Table~\ref{PARAS} gives the hyper-systolic shift constant $K$ as a function of
the number of processors and the number of shifts for the moving array \bm{x}.
\begin{table}[htb]
\begin{center}
\begin{tabular}{|l| ccccccc |}
\hline
p&$16$&$32$&$64$&$128$&$256$&$512$&$1024$\\
\hline
K&2     &4     &6     &8      &12     &16     &24       \\
k&4     &7     &11    &15     &23     &31     &47       \\
\hline
\end{tabular}
\end{center}
\caption[dummy]{Shift constants $K$ for the regular shift sequence as function of 
the number of processors.\label{PARAS}}
\end{table}

\begin{figure}[p]
\epsfxsize=8cm
\centerline{\epsfbox{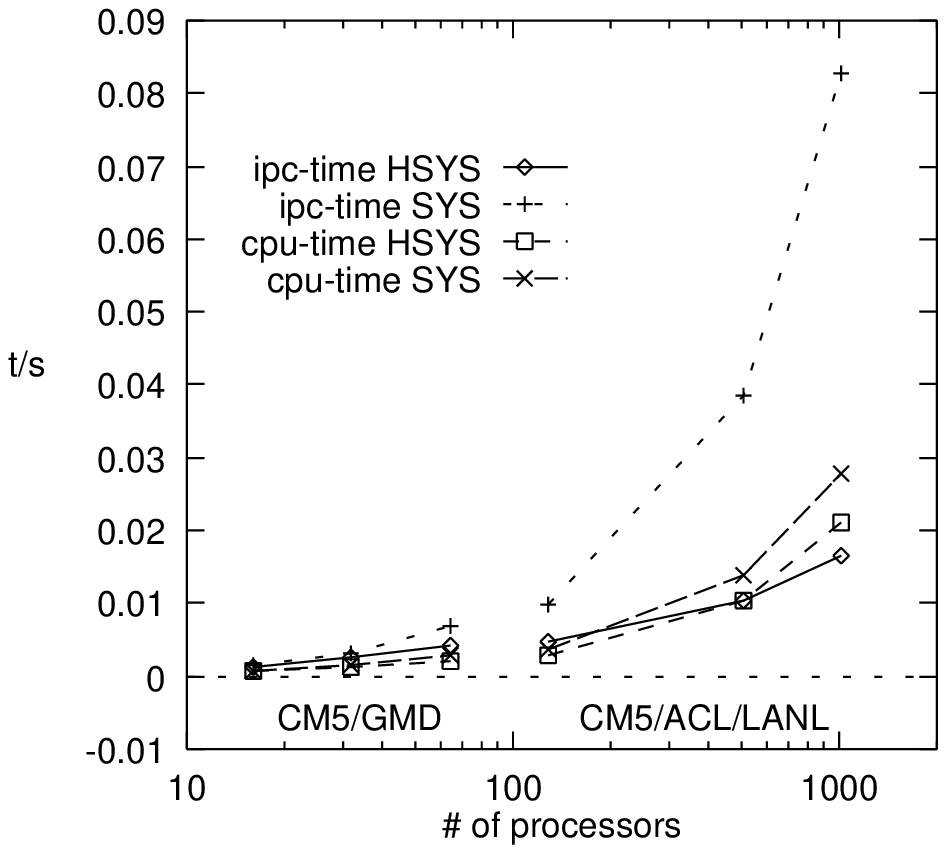}}
\centerline{\epsfbox{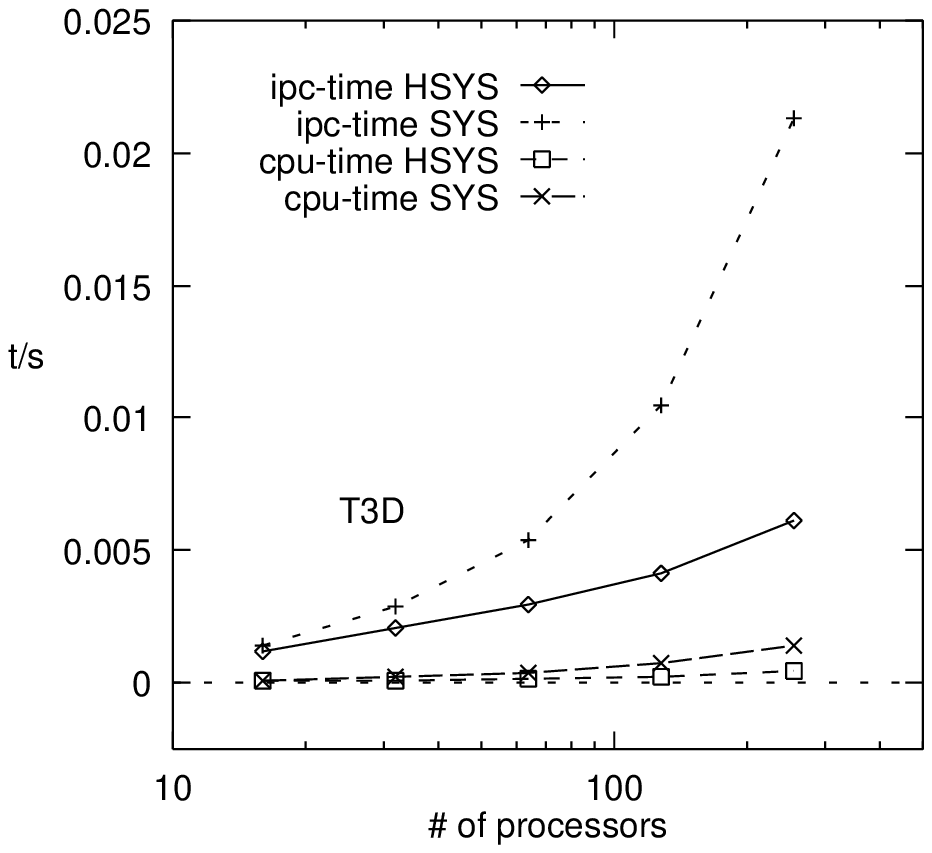}}
\caption[]{
Interprocessor communication time (ipc) and cpu-time for the computation of
\eq{NEWTON} in one molecular dynamics step as function of the number of
processors $p=n$. The lines connect results from different partitions
of one and the same machine.
\label{TIMES}}
\end{figure}

\begin{figure}[p]
\epsfxsize=8cm
\centerline{\epsfbox{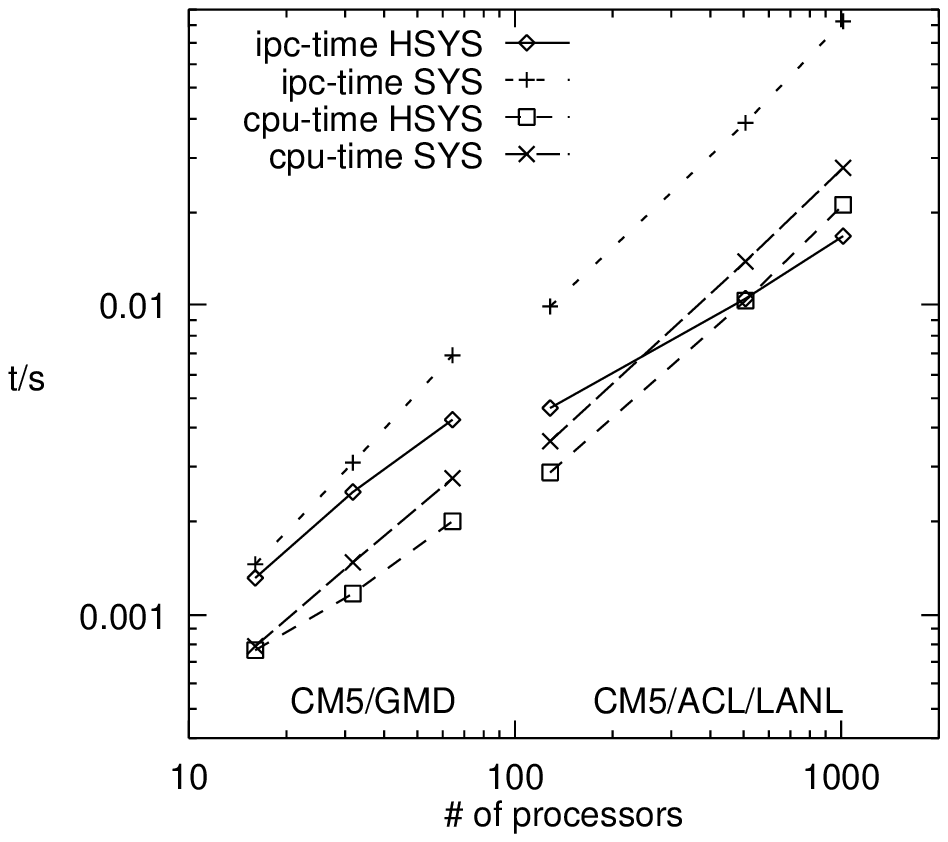}}
\centerline{\epsfbox{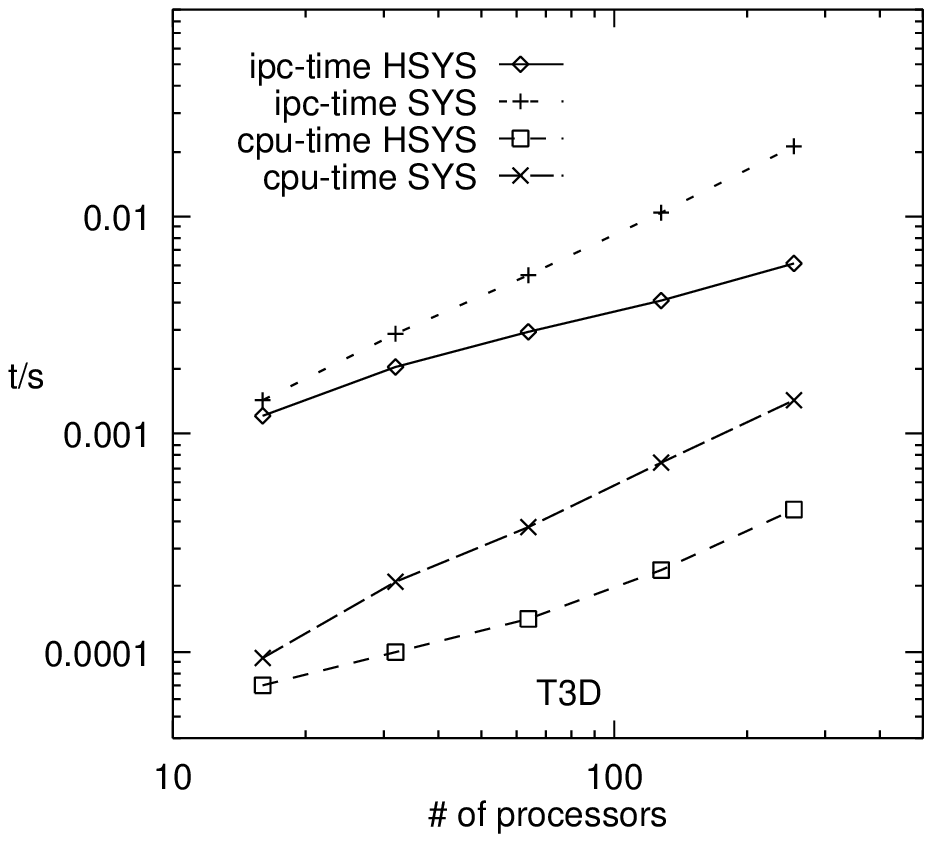}}
\caption[]{
Same plot as before using a logarithmic y-scale.
\label{TIMES_LOG}}
\end{figure}

We plotted the actual interprocessor communication time (ipc) and the
cpu-time for the computation of \eq{NEWTON} in one molecular dynamics
step as function of the number of processors $p=n$ in Fig.~\ref{TIMES},
both for the standard-systolic and the hyper-systolic algorithm.
As indicated, the lines connect results from different partitions of
one and the same machine.  Fig.~\ref{TIMES_LOG} is a logarithmic
replot of Fig.~\ref{TIMES} to exhibit scaling behaviour.  The figure
shows equal scaling of all curves except the curve showing the time
for the hyper-systolic communication.
We note that the hyper-systolic computation appears to need slightly
less cpu-time than the systolic computation.  This is probably due to
a slightly improved pipelining of the cpu-operations, see
footnote~\ref{FOOT:PIPE} on page~\pageref{PIPELINING}.

It turns out that the standard-systolic computation---for $p=n$---is
dominated by interprocessor communication on both the connection
machines CM5 and the CRAY T3D\footnote{Note that we have chosen a
realistic setting of the gravitational force, leading to the compute
intensive calculation of a square root for every coordinate
difference.  In our test model, the system is not given a boundary
condition, and we have cut off the potential for very small
separations of particles, as is necessary in realistic simulations.}.
On the T3D, the effect of the hyper-systolic computation being less
cpu-time intensive is much pronounced due to the time consuming
caching. Most of the time is spent with load and store operations
between cache and local memory.

\begin{figure}[p]
\epsfxsize=8cm
\centerline{\epsfbox{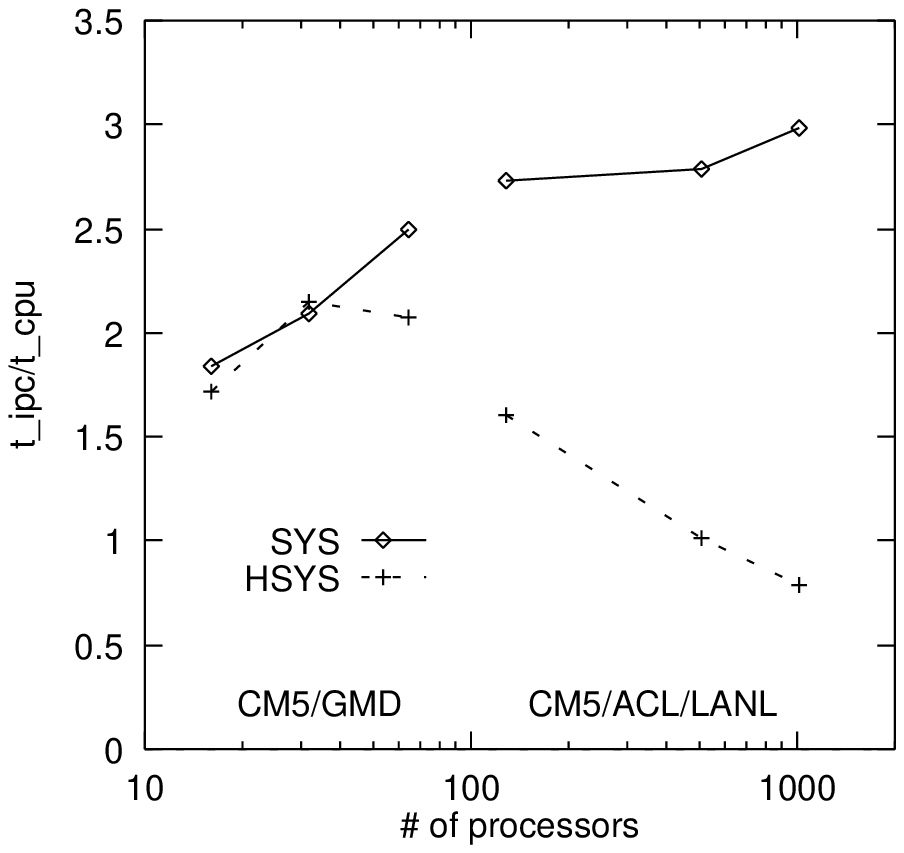}}
\centerline{\epsfbox{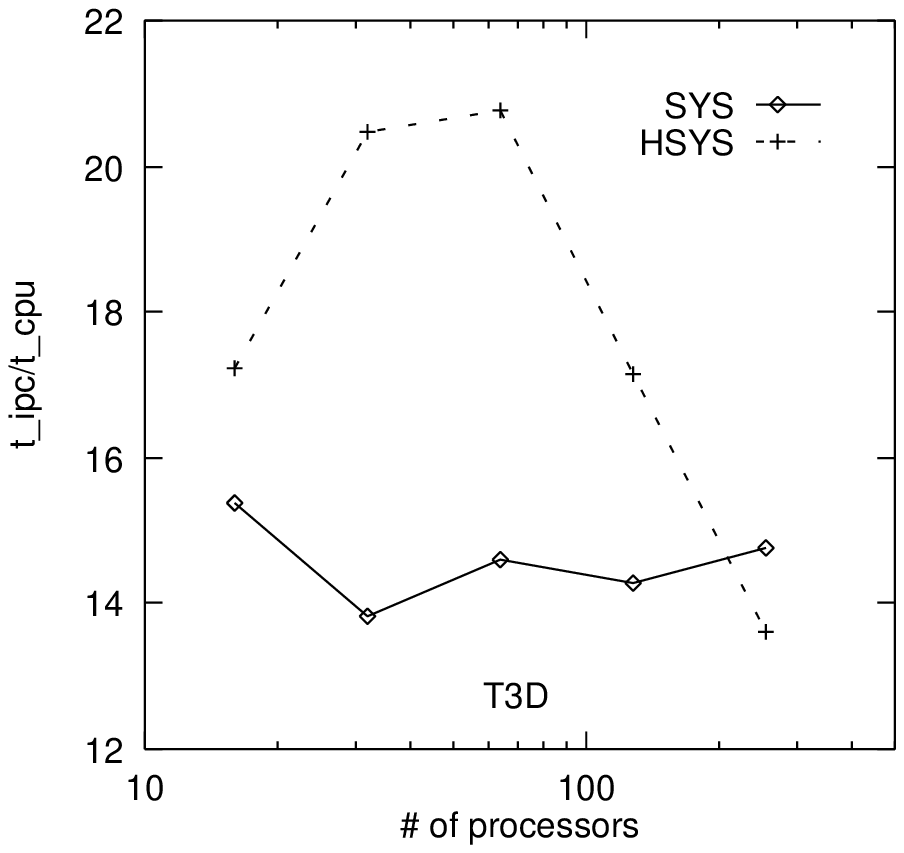}}
\caption[]{
Ratio of interprocessor communication time to cpu-time for the
stand\-ard-systolic and hyper-systolic computation. 
\label{CPU-IPC}}
\end{figure}

Since both, the CPU-time required and the interprocessor communication
time should grow proportional to $n^2$, the ratio between both values
would be expected to vary only slightly for the standard-systolic
method as is the case for the T3D. On the CM5, we have to take into
account logarithmic corrections in interprocessor communication as
explained later.  The result is depicted in Fig.~\ref{CPU-IPC}.
We see that the interprocessor communication will become irrelevant on
massively parallel machines, where the hyper-systolic computation of
\eq{NEWTON} can show its full power. 

\begin{figure}[p]
\epsfxsize=8cm
\centerline{\epsfbox{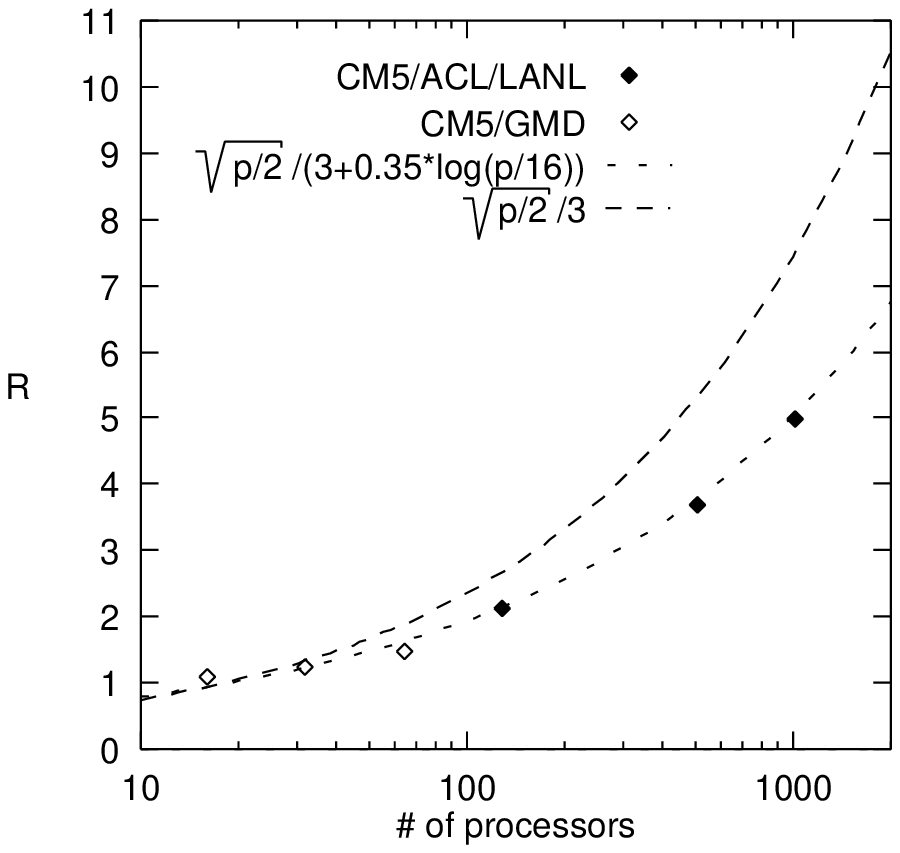}}
\centerline{\epsfbox{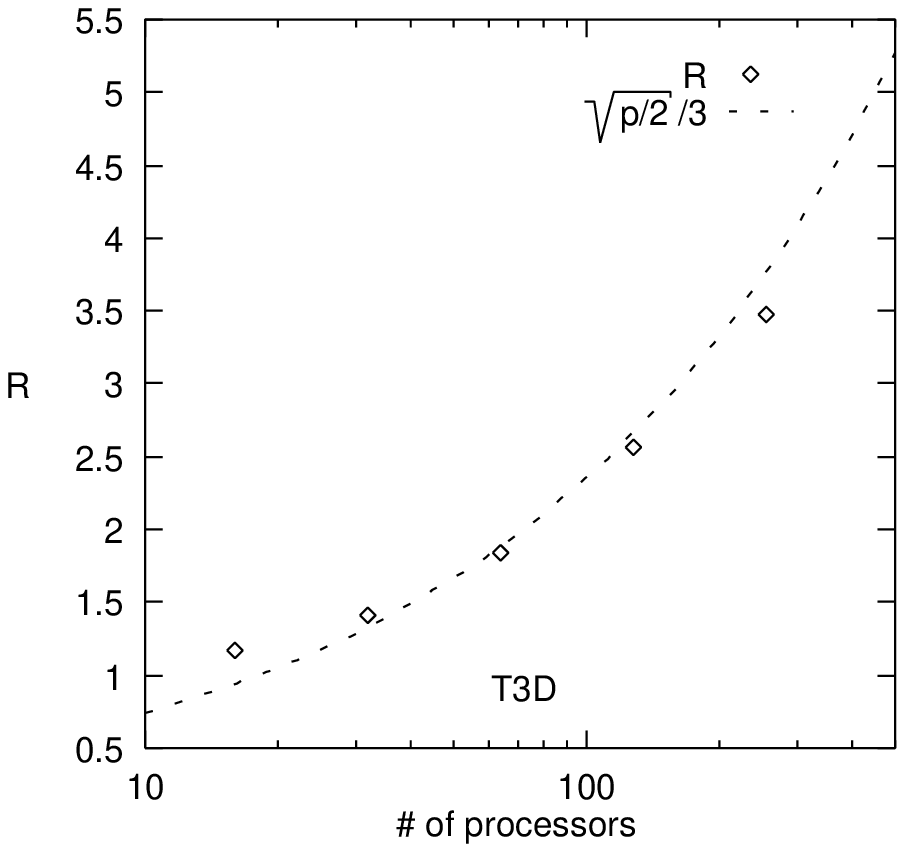}}
\caption[]{
Gain-factor $R$ between the standard-systolic and hyper-systolic
interprocessor communication times. Two theoretically predicted curves
are plotted. On the CM5, we take into account logarithmic corrections
to $\sqrt{p/2}/3$.
\label{CPU-CPU}}
\end{figure}

In Fig.~\ref{CPU-CPU}, we present the gain-factor $R$ between the
standard-systolic and hyper-systolic interprocessor communication
times along with the theoretical prediction.
In our particular implementation, we expect a scaling of the
gain-factor according to $R=\sqrt{p/2}/3$ in contrast to \eq{RATIO}.
The dashed curve represents this naive expectation.  On the CM5,
however, we have to take into account logarithmic corrections to the
scaling law. This is due to the fat-tree hierarchy of its
communication network: communication times to processors in a
`distance' $r$ increase logarithmically with $r$.  Our findings
demonstrate that the hyper-systolic algorithm will at any rate be
superior to conventional methods: the interprocessor communication
expense grows with $n^{\frac{3}{2}}$ as compared to $n^2$.  The full
power of the method can be exploited on big massively parallel
machines.

The additional amount of memory needed in hyper-systolic computations
is tolerable in view of the present machine's and problem's sizes:
even on the largest CM5 with $p=1024$ we only need $2 \times 24 \times
1024$ words of additional memory. Of course, this has to be multiplied
by $n/p$ if we go to $n>p$.

\section{Discussion\label{CONCLUSIONS}}

\paragraph{Summary} We have presented a new parallel algorithm for 
the efficient computation of systems of $n$ elements with pairwise
mutual interactions among all elements, denoted as hyper-systolic
algorithm.  We have developed the hyper-systolic array computation by
generalizing the ordinary systolic realization of the computational
problem.  We illustrated the hyper-systolic algorithm in form of the
canonical abstract automaton representation as it is known from
systolic array computations.

In the context of Additive Number Theory, the hyper-systolic method
very naturally can be formulated as an $h$-range problem $n(h,A_h)$.
We have discussed a solution of the latter problem which is well
suited for the explicit parallel implementation of a hyper-systolic
computation on SIMD and MIMD machines and nearly optimal with respect
to the efficiency of the interprocessor communication.

We applied our method to the computation of the molecular dynamics
evolution of a system of $n$ particles with long range forces. In the
practical implementation on $16$ up to $1024$-node connection machines
CM5 and on $16$ up to $256$-node partitions of the CRAY T3D we could
verify the expected theoretical improvement in interprocessor
communication as compared to the standard-systolic realization.

\paragraph{Conclusions} The hyper-systolic way of parallel 
computing can be extended to a variety of parallel computations which
are usually done by systolic $n^2$-loops.  The potential of the method
appears of great promise. It might even affect the very architectural
issue of parallel machines, as it is efficiently applicable on
massively parallel machines with simple (and therefore cheap) toroidal
next-neighbour communication networks \cite{QUADRICS}.  As an example,
we will implement the hyper-systolic array computation on Quadrics
(APE 100) machines produced by Alenia Spazio S.p.A.\
\cite{QUADRICS}. These machines are equipped with a simple
three-dimensional next-neighbour network. We expect the hyper-systolic
computation to speed-up the Quadrics performance considerably
\cite{HOEBER}.  The full payoff of the idea will be achieved, of
course, on real massively parallel platforms such as described in
\cite{CHRIST}.

We hope that the hyper-systolic approach of tackling $n^2$-loop
problems will provide an essential improvement for practical
simulations.

\section*{Acknowledgements}

We are indebted to Prof.\ Dr.\ Harald Scheid, Mathematics Department,
University of Wuppertal, and Prof.\ Dr.\ Gerd Hof\-mei\-ster,
Institute for Additive Number Theory, University of Mainz, for
illuminating discussions as to Additive Number Theory and the
classification of hyper-systolic algorithms.

Part of our numerical results has been achieved on the 1024-node connection
machine CM5 of the Advanced Computing Laboaratory at Los Alamos National
Laboratory, USA.  We are grateful to the staff of ACL and in particular to
Dr.\ Daryl W. Grunau ACL/LANL/TMC.  We thank Dr.\ Peter Ossadnik for his
support in testing the hyper-systolic code on the connection machines CM5 at
GMD/HLRZ, Germany and for his useful hints as to our work.

A.\ B.\ thanks Prof.\ Dr.\ R.\ Kenway from EPCC, United Kingdom, for
providing access to the Cray T3D at EPCC.  We acknowledge Dr.\ P.\
Ueberholz for his support at the EPCC T3D site.

Th.\ L.\ and K.\ S.\ thank Prof.\ Dr.\ Nikolay Petkov, University of
Groningen, for initiating us into the art of systolic computing.  

The work of Th.\ L.\ is supported by the Deutsche
Forschungsgemeinschaft DFG under grant No.\ Schi 257/5-1.

\end{document}